\documentclass[prd,preprint,nofootinbib]{revtex4-1} 

\usepackage{latexsym,verbatim,enumerate}
\usepackage{amssymb}
\usepackage{amsfonts}
\usepackage{amsmath,color}
\usepackage{graphicx}
\usepackage{bm}
\usepackage[utf8]{inputenc}
\usepackage[T1]{fontenc}

\usepackage{hyperref}

\hypersetup{
    colorlinks=true,
    linkcolor=red,
    citecolor=red,
    filecolor=red,      
    urlcolor=red,
    pdftitle={Gravitomagnetic Frame Memory},
    pdfpagemode=FullScreen,}

\def\mb#1{\mathbf{#1}}

\def\ber{\begin{eqnarray}}
\def\eer{\end{eqnarray}}
\def\beq{\begin{equation}}
\def\eeq{\end{equation}}

\def\rmd{{\rm d}}

\def\ed{\end{document}}

\def\dT#1{\frac{\mathrm{d} #1}{\mathrm{d}T}}

\def\sT{\sin \left(\omega T \right)}
\def\cT{\cos \left(\omega T \right)}

\makeatletter
\let\jnl@style=\rm
\def\ref@jnl#1{{\jnl@style#1}}

\def\aj{\ref@jnl{AJ}}                   
\def\actaa{\ref@jnl{Acta Astron.}}      
\def\araa{\ref@jnl{ARA\&A}}             
\def\apj{\ref@jnl{ApJ}}                 
\def\apjl{\ref@jnl{ApJ}}                
\def\apjs{\ref@jnl{ApJS}}               
\def\ao{\ref@jnl{Appl.~Opt.}}           
\def\apss{\ref@jnl{Ap\&SS}}             
\def\aap{\ref@jnl{A\&A}}                
\def\aapr{\ref@jnl{A\&A~Rev.}}          
\def\aaps{\ref@jnl{A\&AS}}              
\def\azh{\ref@jnl{AZh}}                 
\def\baas{\ref@jnl{BAAS}}               
\def\bac{\ref@jnl{Bull. astr. Inst. Czechosl.}}
\def\caa{\ref@jnl{Chinese Astron. Astrophys.}}
\def\cjaa{\ref@jnl{Chinese J. Astron. Astrophys.}}
\def\icarus{\ref@jnl{Icarus}}           
\def\jcap{\ref@jnl{J. Cosmology Astropart. Phys.}}
\def\jrasc{\ref@jnl{JRASC}}             
\def\memras{\ref@jnl{MmRAS}}            
\def\mnras{\ref@jnl{MNRAS}}             
\def\na{\ref@jnl{New A}}                
\def\nar{\ref@jnl{New A Rev.}}          
\def\pra{\ref@jnl{Phys.~Rev.~A}}        
\def\prb{\ref@jnl{Phys.~Rev.~B}}        
\def\prc{\ref@jnl{Phys.~Rev.~C}}        
\def\prd{\ref@jnl{Phys.~Rev.~D}}        
\def\pre{\ref@jnl{Phys.~Rev.~E}}        
\def\prl{\ref@jnl{Phys.~Rev.~Lett.}}    
\def\pasa{\ref@jnl{PASA}}               
\def\pasp{\ref@jnl{PASP}}               
\def\pasj{\ref@jnl{PASJ}}               
\def\rmxaa{\ref@jnl{Rev. Mexicana Astron. Astrofis.}}%
\def\qjras{\ref@jnl{QJRAS}}             
\def\skytel{\ref@jnl{S\&T}}             
\def\solphys{\ref@jnl{Sol.~Phys.}}      
\def\sovast{\ref@jnl{Soviet~Ast.}}      
\def\ssr{\ref@jnl{Space~Sci.~Rev.}}     
\def\zap{\ref@jnl{ZAp}}                 
\def\nat{\ref@jnl{Nature}}              
\def\iaucirc{\ref@jnl{IAU~Circ.}}       
\def\aplett{\ref@jnl{Astrophys.~Lett.}} 
\def\apspr{\ref@jnl{Astrophys.~Space~Phys.~Res.}}
\def\bain{\ref@jnl{Bull.~Astron.~Inst.~Netherlands}}
\def\fcp{\ref@jnl{Fund.~Cosmic~Phys.}}  
\def\gca{\ref@jnl{Geochim.~Cosmochim.~Acta}}   
\def\grl{\ref@jnl{Geophys.~Res.~Lett.}} 
\def\jcp{\ref@jnl{J.~Chem.~Phys.}}      
\def\jgr{\ref@jnl{J.~Geophys.~Res.}}    
\def\jqsrt{\ref@jnl{J.~Quant.~Spec.~Radiat.~Transf.}}
\def\memsai{\ref@jnl{Mem.~Soc.~Astron.~Italiana}}
\def\nphysa{\ref@jnl{Nucl.~Phys.~A}}   
\def\physrep{\ref@jnl{Phys.~Rep.}}   
\def\physscr{\ref@jnl{Phys.~Scr}}   
\def\planss{\ref@jnl{Planet.~Space~Sci.}}   
\def\procspie{\ref@jnl{Proc.~SPIE}}   

\makeatother

\begin{document}

\author{Matteo Luca Ruggiero}
\email{matteoluca.ruggiero@unito.it}
\affiliation{Dipartimento di Matematica ``G.Peano'', Universit\`a degli studi di Torino, Via Carlo Alberto 10, 10123 Torino, Italy}
\affiliation{INFN - Sezione di Torino, Via Pietro Giuria 1, 1025 Torino, Italy}
\author{Davide Astesiano}
\email{davide.astesiano@venturilab.ch}
\affiliation{Science Institute, University of Iceland,
Dunhaga 3, 107, Reykjav\'{\i}k, Iceland} 
\affiliation{Mathematics Division, Venturi Lab SA, Route du Pâqui 1, 1720 Corminboeuf, Switzerland}

\date{\today}

\title{Gravitomagnetic Frame Memory}

\begin{abstract}
We investigate the response of spatially separated spinning test particles to a finite-duration plane gravitational wave in the proper detector frame,  {working to linear order in the wave amplitude.}
 Using Fermi normal coordinates and the gravitoelectromagnetic decomposition of the local metric, the spin evolution is governed, at leading order in the distance from the reference world-line, by the magnetic part of the Riemann tensor. A finite curvature profile with a non-vanishing gravitomagnetic pulse area induces a residual relative rotation of the spins,
proportional to the time integral of the local gravitomagnetic field.
The resulting change persists after the curvature has vanished and may be
interpreted as a local frame-memory observable. This effect has the structure
of a velocity-type memory of finite plane-wave curvature profiles and provides
a direct operational probe of the integrated magnetic part of spacetime
curvature.
 \end{abstract}

\maketitle

\section{Introduction} 
\label{sec:intro}

Gravitational-wave memory refers to a persistent change in a detector
observable after the radiative field has passed. The best-known example is
displacement memory, in which initially freely falling test bodies acquire a
permanent change in their relative separation \cite{Zhang:2017geq}. 

Gravitational memory is now known to encompass a broad variety of physical manifestation \cite{Zeldovich:1974gvh,braginsky1987gravitational},
these include kick, or velocity, memory generated by plane gravitational waves
\cite{Chakraborty:2022qvv} and by compact astrophysical sources
\cite{Divakarla}, spin memory observable through
relative time delays in Sagnac interferometers
\cite{Pasterski:2015tva}.
Much of this diverse phenomenology can be understood within the common
framework of asymptotic symmetries
\cite{Strominger:2017zoo}. In particular, these effects
are closely connected with the Bondi--Metzner--Sachs (BMS) symmetries of
asymptotically flat gravity
\cite{Bondi1962}.

In this work we study the interaction of gravitational waves with spinning systems, which can be studied
particularly transparently in the proper detector frame, where the spacetime metric is expressed in  terms of Fermi coordinates; the latter form a quasi-Cartesian coordinate system that can be constructed in a neighborhood of a given observer. Fermi coordinates provide a geometrically defined coordinate system adapted
to the world-line of a reference observer and to an orthonormal tetrad
transported along it \cite{MTW}. They have a direct operational meaning,
since the corresponding spatial coordinates are defined through spacelike
geodesics orthogonal to the reference world-line \cite{synge1960relativity}. In this sense, they provide
a natural local framework for describing measurements performed in the
neighborhood of the observer. In particular, these coordinates provide a local and operational description of the gravitational
field in terms of the electric and magnetic parts of the Riemann tensor. In
this framework, the magnetic part of the wave produces a precession of a
spinning particle, 
and may also give rise to gravitomagnetic resonance
\cite{RuggieroOrtolan2020}.
Since the local gravitomagnetic field
vanishes on the reference geodesic and grows linearly with the spatial
distance from it, the physically relevant quantity is the relative
orientation of spins or local tetrads situated at different positions in the
Fermi frame.

A related but distinct phenomenon is the gyroscopic gravitational memory
introduced in Refs.~\cite{Seraj:2022qyt,SerajOblak2023}. There, a gyroscope
located far from an isolated source is compared with an optical frame tied to
distant stars. Its accumulated precession contains a linear contribution
associated with spin memory and a nonlinear term that measures the net
helicity of the gravitational-wave burst. The corresponding precession rate
is subleading in the large-distance expansion and depends on the angular
structure of the radiation at null infinity. Gyroscopic memory has also been
evaluated explicitly for quasi-circular compact binaries
\cite{Faye:2024utu}. The effect discussed here should therefore not be identified with the
asymptotic spin-memory observable. Rather, it has a different, genuinely
local origin.  {Other  studies \cite{revpaper1,revpaper2} have investigated persistent gravitational-wave effects on particle trajectories, spin transport, and wave-packet propagation; here, we consider a complementary configuration in which spins are held at fixed spatial positions.\\
\indent In particular, working to linear order in the gravitational wave amplitude  and adopting some simplifying assumptions on the geometry of the system, introduced to highlight the physical origin of the effect, we compare the spin carried by a test body at a finite spatial
separation from the reference observer with a reference spin at the origin
of the freely falling, non-rotating Fermi frame associated with the given observer}. The resulting observable is
a relative-frame observable: it measures the net rotation accumulated
between spatially separated local frames when the time integral of the
magnetic curvature does not vanish. For a finite plane-wave curvature
profile, a nonzero gravitomagnetic pulse area therefore leaves a persistent
change in their relative orientation after the curvature has vanished.
This provides a simple local realization of frame memory, closely related
to the velocity-memory properties of sandwich plane waves \cite{Zhang:2017geq}.


\section{The Local Metric}\label{sec:local}

The spacetime metric in the laboratory frame is expressed in terms of Fermi coordinates, which, as discussed above, are constructed from scalar invariants and therefore provide a natural framework for defining the measurement process in terms of observable quantities. This metric depends both on how the laboratory frame moves in spacetime (i.e. on its acceleration and the rotation of its axes) and on the spacetime curvature through the Riemann  tensor \citep{Ni:1978di,Li:1979bz,1982NCimB..71...37F,marzlin}. In particular, we are interested in the effects of a gravitational wave which are described by the Riemann tensor. Accordingly, by exploiting the gravitoelectromagnetic analogy discussed by  \citet{Ruggiero_2020}, using coordinates $(cT,X,Y,Z)$, for a laboratory in free-fall and non rotating, up to quadratic displacements $|X^{i}|$ from the reference world-line, the metric can be written in the form\footnote{Latin indices refer to space coordinates, while Greek indices to spacetime ones. Moreover,  we will use bold-face symbols like  $\mb W$ to refer to vectors in the Fermi frame. The spacetime signature is $(-1,1,1,1)$.}
\beq
\mathrm{d} s^2= -c^2 \left(1-2\frac{\Phi}{c^2}\right)\rmd T^2 -\frac4c \mathcal A_{i}\rmd X^{i}\rmd T  +
 \left(\delta_{ij}+2\frac{\Psi_{ij}}{c^2}\right)\rmd X^i \rmd X^j\ , \label{eq:weakfieldmetric11}
\eeq
with
\begin{eqnarray}
\Phi (T, { X^{i}})&=&-\frac{c^{2}}{2}R_{0i0j}(T )X^iX^j, \label{eq:defPhiG}\\
\mathcal A^{}_{i}(T ,{X^{i}})&=&\frac{c^{2}}{3}R_{0jik}(T )X^jX^k, \label{eq:defAG}\\
\Psi_{ij} (T, {X^{i}}) & = & -\frac{c^{2}}{6}R_{ikjl}(T)X^{k}X^{l}, \label{eq:defPsiG}
\end{eqnarray}
where $\Phi$ and $\mathcal A_{i}$ are, respectively, the {gravitoelectric} and {gravitomagnetic} potential, and $\Psi_{ij}$ is the perturbation of the spatial metric \cite{manasse1963fermi,MTW,Mashhoon:2003ax,Ruggiero_2020,Ruggiero:2021uag,Ruggiero:2023ker}. In the above equations $R_{\alpha \beta \gamma \delta}(T)$ is the projection of the Riemann tensor on the tetrad attached to the laboratory frame, and it is evaluated  along the reference world-line. We are interested in the laboratory metric describing the field of a plane gravitational wave propagating in the $X$ direction with frequency $\omega$.  In this case, the components of the gravitomagnetic field  $\displaystyle \mb B= \bm \nabla \wedge \mb{\mathcal A}$  turn out to be (see e.g. \citet{Ruggiero_2020})
\begin{eqnarray}
B^{}_{X}(T)  &=& 0, \label{eq:BX} \\
 B^{}_{Y}(T)  &=& -\frac{\omega^{2}}{2}\left[-A^{\times} \cT Y+A^{+} \sT Z \right]  \label{eq:BY}, \\
  B^{}_{Z}(T)  &=& -\frac{\omega^{2}}{2}\left[A^{+}\sT Y+A^{\times}\cT Z \right] \label{eq:BZ}.
\end{eqnarray}
where  $A^{+},A^{\times}$ are the amplitude of wave in the two polarization states. We point out that the expression of the metric tensor used to obtain the gravitomagnetic field is valid in the large wavelength limit, which means that the typical dimension $\ell$ of the laboratory frame  is negligible with respect to the wavelength $\lambda$. More accurate expressions can be obtained, which contain higher order terms in the small parameter $\displaystyle \epsilon = \frac{\ell}{\lambda}$  (see e.g. \citet{Ruggiero:2022gzl} and references therein).  The series expansion can be exactly summed to obtain a compact form \cite{1982NCimB..71...37F,Berlin:2021txa}.

\section{The Effect on Spinning Particles}\label{sec:spin}

 {A classical spinning test particle held at a fixed spatial position in the Fermi frame undergoes, to leading order in the Fermi coordinates and in the gravitational perturbation, an evolution governed by the gravitomagnetic precession equation \cite{Ruggiero_2020,mashhoonentropy}
\beq
\dT{\mb S}= \frac{1}{c}\mb B \times \mb S \label{eq:spinevol11}
\eeq
Keeping the particle fixed generally requires nongravitational forces, which are assumed to be torque-free so that they do not induce any additional spin precession.} Let us clarify the interpretation of  Eq.(\ref{eq:spinevol11}). The reference spin at the origin is understood as an ideal gyroscopic direction associated with the parallel-transported tetrad along the reference geodesic defining the Fermi frame: accordingly, its components remain constant in that frame, even in the presence of non-vanishing curvature. The latter manifests itself through tidal effects away from the reference world-line: in the Fermi frame the local gravitomagnetic field  (\ref{eq:BX})-(\ref{eq:BZ}) is curvature-generated and vanishes on the reference geodesic. We remark that the reference world-line is thus used to define the local non-rotating frame against which the off-origin spinning particle is compared,  rather than to represent the full translational dynamics of a finite-spin body governed by the Mathisson-Papapetrou-Dixon equation  \cite{Papapetrou1951,Dixon1970}. The off-origin spin, held at fixed Fermi coordinates, instead evolves according to the gravitomagnetic precession law (\ref{eq:spinevol11}): the resulting spin rotation is therefore measured \textit{relative} to the reference spin at the origin.

As we said,   the reference spin
$\mathbf S_{\rm ref}$ remains unchanged. Initially,
\begin{equation}
\Delta\mathbf S_{\rm rel}(0)
\equiv
\mathbf S(0)-\mathbf S_{\rm ref}(0)
=0.
\end{equation}
After the gravitational-wave burst has passed,
\begin{equation}
\Delta\mathbf S_{\rm rel}(\tau)
\equiv
\mathbf S(\tau)-\mathbf S_{\rm ref}(\tau)
\end{equation}
which is generically non-vanishing.

 {Let us define the {gravitomagnetic pulse-area vector} as
\begin{equation}
\boldsymbol{\Theta}_{G}=
\frac{1}{c}
\int_{-\infty}^{+\infty}\mathbf B(T)\,dT ,
\label{eq:thetaG}
\end{equation}
where $\mathbf B(T)$ is the local gravitomagnetic field measured in the Fermi frame. The quantity $\boldsymbol{\Theta}_{G}$ is a vector whose direction is determined by the time-integrated orientation of the gravitomagnetic field, while its magnitude measures the total accumulated gravitomagnetic rotation produced by the burst. The residual spin rotation is therefore controlled by the component of $\boldsymbol{\Theta}_{G}$ transverse to the initial spin direction. In fact, to linear order in the gravitational-wave amplitude, the result can be
written in the general form
\begin{equation}
\Delta\mathbf S_{\rm rel}
=
\frac{1}{c}
\left[
\int_{-\infty}^{+\infty}
\mathbf B(T)\,dT
\right]
\times\mathbf S^0
=
\boldsymbol{\Theta}_{G}\times\mathbf S^0 ,
\label{eq:spinmemory}
\end{equation}
which is independent of the choice of the initial spin orientation and on its location, and the relevant quantities are expressed as vectors in the Fermi frame.}



Equation (\ref{eq:spinmemory}) makes clear that the residual effect is not
determined by the instantaneous value of the gravitomagnetic field, but by
its integrated pulse area. In this respect, the result is reminiscent of
magnetic-resonance dynamics, where the net rotation induced by a pulse is
controlled by the time-integrated driving field \cite{levitt2001spin}.

The physical character of this effect can be stated directly in geometrical
terms. The gravitomagnetic field entering Eq.~(\ref{eq:spinmemory}) is
constructed from tetrad projections of the Riemann tensor evaluated along
the reference world-line. At linear order in the gravitational-wave
perturbation, these curvature components are gauge invariant. Moreover, the
observable is the relative orientation of spins, or equivalently of local
spatial frames, defined with respect to the Fermi-Walker transported tetrad
of the reference observer. The residual rotation is therefore not a
coordinate artifact of the metric perturbation, but a local observable
defined through spacetime curvature and relative frame orientation.

\section{Comments}\label{sec:com}

The result obtained above has the characteristic structure of a memory
observable. Consider a finite curvature profile such that the gravitational
field vanishes in both the early- and late-time regions. A detector observable
$\mathcal O$ exhibits memory when
\[
\Delta \mathcal O
\equiv
\lim_{t\to +\infty}\mathcal O(t)
-
\lim_{t\to -\infty}\mathcal O(t)
\neq 0 .
\]

even though the radiative curvature has vanished in both asymptotic regimes.

 {In the present case we  define
\begin{equation}
\Delta\mathbf S_{\rm mem}
\equiv
\lim_{T\rightarrow+\infty}
\Delta\mathbf S_{\rm rel}(T)
-
\lim_{T\rightarrow-\infty}
\Delta\mathbf S_{\rm rel}(T).
\label{eq:memoryobservable}
\end{equation}
For initially aligned spins, $\Delta\mathbf S_{\rm rel}(-\infty)=0$ and, to linear order in the gravitational-wave amplitude,  from Eq. (\ref{eq:spinmemory}) we have
\begin{equation}
\Delta\mathbf S_{\rm mem}
=
\frac{1}{c}
\left[
\int_{-\infty}^{+\infty}
\mathbf B(T)\,dT
\right]
\times\mathbf S^0 =\boldsymbol{\Theta}_{G}\times\mathbf S^0 \label{eq:spinmemory2}
\end{equation}
Consequently, a finite curvature profile produces a non-vanishing local
frame memory if and only if its gravitomagnetic pulse area $\boldsymbol{\Theta}_{G}$, projected
transversely to the initial spin direction, does not vanish. Since Eq. (\ref{eq:spinmemory2}) is expressed in vector form in the Fermi frame, the law itself is not tied to the particular choices of spin orientation and particle position adopted above; these choices only determine the components of $\mathbf S^0$ and $\boldsymbol{\Theta}_{G}$, and hence the magnitude and direction of the resulting memory.}

A classical spin acts as a gyroscope and therefore defines a local
non-rotating spatial direction. Equation~(\ref{eq:spinmemory2}) can thus be
equivalently interpreted as describing a residual relative rotation between
the local spatial frames associated with two separated observers. Existing
gravitational-memory observables are commonly formulated in terms of
persistent changes in relative displacement, relative velocity, or timing.
The observable considered here is instead encoded in the residual relative
orientation of local frames. Notice that the magnitude of each spin remains unchanged throughout the
process. The memory therefore concerns the orientation of the angular-momentum
direction rather than its magnitude.

For the standard displacement memory, the relevant detector observable is
the relative separation $\xi^i$ of freely falling test masses. In the
weak-field, slow-motion and radiation-zone regime, the geodesic-deviation
equation has the schematic form $\ddot{\xi}^{\,i} \sim R^{i}{}_{0j0}\,\xi^j$
while the radiative curvature generated by a quadrupolar source scales as $ R^{\rm rad}
\propto Q^{(4)}$.
Two time integrations therefore relate the ordinary displacement-memory
observable to the asymptotic change of the second derivative of the source
quadrupole $ \Delta\xi
\propto
\Delta Q^{(2)}$. 
For standard asymptotically free massive sources,
\begin{equation}
Q^{(2)}_{ij}
\sim
\sum_k m_k v_k^i v_k^j ,
\end{equation}
which gives the usual ordinary displacement memory \cite{bieri}.

The frame observable considered here has a different derivative structure.
From Eq.~(\ref{eq:spinmemory}),
\begin{equation}
\Delta\mathbf S_{\rm mem}
\propto
\int R^{\rm rad}\,dT ,
\end{equation}
so that, at the level of the radiative quadrupole expansion,
\begin{equation}
\Delta\mathbf S_{\rm mem}
\propto
\int Q^{(4)}\,dT
\propto
\Delta Q^{(3)} .
\label{eq:q3memory}
\end{equation}
Thus, in contrast with ordinary displacement memory, the present observable
contains one rather than two time integrations of the radiative curvature.
In this sense it has the derivative structure characteristic of a
velocity-type memory.

For asymptotically free massive sources this contribution vanishes, since
\begin{equation}
Q^{(3)}_{ij}\longrightarrow 0
\qquad
(T\rightarrow\pm\infty),
\end{equation}
when the bodies approach uniform rectilinear motion in the remote past and
future. Hence the source conditions that generate the standard ordinary
displacement memory do not, by themselves, generate the local frame-memory
observable considered here.

A qualitatively different situation may arise for an oscillating source
undergoing a sufficiently rapid transition between two dynamical regimes.
For a quasi-circular binary with reduced mass $\mu$, orbital separation $d$,
and orbital frequency $\omega$, the characteristic amplitude of the third
time derivative of the mass quadrupole scales as
\begin{equation}
|Q^{(3)}|
\sim
\mu d^2\omega^3 .
\label{eq:q3scale}
\end{equation}

For a strictly stationary periodic orbit, the corresponding radiative
contribution is oscillatory and does not define an asymptotic jump of
$Q^{(3)}_{ij}$. However, if the source undergoes a non-adiabatic transition
on a timescale comparable to or shorter than the orbital period, the
cancellation of the oscillatory contribution over the transition need not
be complete. The characteristic scale associated with a change between two
orbital configurations is then
\begin{equation}
|\delta Q^{(3)}|
\sim
\mu
\left|
d_+^2\omega_+^3
-
d_-^2\omega_-^3
\right|,
\label{eq:q3jump}
\end{equation}
where the signs $-$ and $+$ denote characteristic values before and after
the transition.

Equation~(\ref{eq:q3jump}) should be understood as an estimate of the
available amplitude scale rather than as a universal asymptotic jump:
the actual integrated response also depends on the orbital phase and on the
time profile of the transition. A detailed source model is therefore
required to determine whether a given binary transition produces a
non-vanishing gravitomagnetic pulse area.

As for the possible observability of the effect, the extremely small
amplitudes of gravitational waves imply that the rotation predicted for an
individual microscopic spin is correspondingly tiny. It is nevertheless
natural to ask whether a collective spin system could provide a macroscopic
transduction mechanism. Ferromagnetic media have previously been considered
in connection with gravitational-wave-induced spin dynamics  \cite{RuggieroOrtolan2020},
and they suggest an intriguing possible setting also for the present effect.

However, a ferromagnetic material cannot be regarded simply as an ensemble
of independent freely precessing gyroscopes: exchange interactions,
magnetic anisotropy, damping, domain dynamics, and spin--lattice coupling
determine the collective response. Establishing whether the local frame
rotation derived here can be converted into a persistent macroscopic change
of magnetization therefore requires a separate condensed-matter analysis.
We leave this question for future investigation.

In  summary, we remark that the main result of this work is purely geometrical and does not
rely on a specific material realization. A finite gravitational-wave
curvature profile can leave a persistent relative rotation between
spatially separated local inertial frames whenever the integrated magnetic
part of the curvature is non-vanishing. The corresponding observable is
controlled by one time integral of the radiative curvature and therefore
has the derivative structure of a velocity-type memory. It provides a local
operational characterization of the gravitomagnetic content accumulated
during the passage of a gravitational-wave burst.

\section*{Acknowledgements}

MLR acknowledges the support of the Istituto Italiano di Fisica Nucleare (INFN) via the iniziativa specifica QGSKY and thanks the Gruppo Nazionale per la Fisica Matematica (GNFM) for its support.


\bibliography{refs} 

\end{document}